\documentclass[pdftext,a4wide,12pt]{article}
\usepackage[dvips]{graphicx}
\usepackage{epsfig}
\usepackage{amsmath}
\usepackage{amsfonts}
\usepackage{graphics,color}
\def\a{\alpha}

\def\f{\frac}

\def\lf{\left}

\def\nn{\nonumber}

\def\ri{\right}
\def\rar{\rightarrow}

\def\tr{\textrm}

\newcommand{\be}{\begin{equation}}
\newcommand{\ee}{\end{equation}}
\newcommand{\bea}{\begin{eqnarray}}
\newcommand{\eea}{\end{eqnarray}}
\newcommand{\bl}{\begin{align*}}
\newcommand{\el}{\end{align*}}

\newcommand{\vect}[1]{{\overrightarrow{#1}}}

\newcommand{\hba}{\hat{\mathbf{a}}}
\newcommand{\bba}{{\mathbf{a}}}

\newcommand{\zet}{\mathbb{Z}}
\newcommand{\ba}{\begin{array}}
\newcommand{\ea}{\end{array}}

\makeatletter \@addtoreset{equation}{section} \makeatother

\begin{document}
\begin{titlepage}

%\version\versionno

\hbox to \hsize{\hbox{\tt hep-th/0604092}\hss
    \hbox{\small{\tt MCTP-06-05}}}

\vspace{2 cm}

\centerline{\bf \Large Finite Heisenberg Groups and Seiberg Dualities}

\vspace{.6cm}
\centerline{\bf \Large in Quiver Gauge Theories}

%\vspace{.6cm} \centerline{\bf \Large }
\vspace{2 cm}
 \centerline{\large Benjamin A. Burrington, James T. Liu, Manavendra  Mahato }

\vspace{.3cm}
\centerline{\large  and Leopoldo A. Pando Zayas}

\vspace{2 cm}

\centerline{\it Michigan Center for Theoretical Physics}
\centerline{\it Randall Laboratory of Physics, The University of
Michigan} \centerline{\it Ann Arbor, MI 48109--1040}

\vspace{2 cm}

\begin{abstract}
A large class of quiver gauge theories admits the action of finite Heisenberg groups of the form ${\rm Heis}(\mathbb{Z}_q\times \mathbb{Z}_q)$.  This
Heisenberg group is generated by a manifest $\mathbb Z_q$ shift symmetry
acting on the quiver along with a second $\mathbb Z_q$ rephasing (clock)
generator acting on the links of the quiver.  Under Seiberg duality, however,
the action of the shift generator is no longer manifest, as the dualized
node has a different structure from before.  Nevertheless, we demonstrate
that the $\mathbb Z_q$ shift generator acts naturally on the space of all
Seiberg dual phases of a given quiver.  We then prove that the space of
Seiberg dual theories inherits the action of the original finite Heisenberg group,
where now the shift generator $\mathbb{Z}_q$ is a map among fields belonging
to different Seiberg phases.  As examples, we explicitly consider the action
of the Heisenberg group on Seiberg phases for $\mathbb{C}^3/\mathbb{Z}_3$,
$Y^{4,2}$ and $Y^{6,3}$ quiver.
\end{abstract}
\end{titlepage}

%%%%%%%%%%%%%%%%%%%%%%%%%%%%%%%%%%%%%%%%%%%%%%%%%%%%%%%%%%%%%%%%%%%%%%%%%%
\section{Introduction}
%%%%%%%%%%%%%%%%%%%%%%%%%%%%%%%%%%%%%%%%%%%%%%%%%%%%%%%%%%%%%%%%%%%%%%%%%%
Generalizing an observation of \cite{grw}, it was shown in
\cite{Burrington:2006uu}
that for a class of $\mathbb Z^q$ orbifold quiver gauge theories with gauge
group $SU(N)^p$, there is a set of discrete transformations $A$, $B$ and $C$
satisfying
\be
\label{heis}
A^{q}=B^{q}=C^{q}=1, \qquad AB=BAC.
\ee
Here the $A$ generator is inherited from the $\mathbb Z^q$ orbifold action,
and corresponds to the manifest $\mathbb Z^q$ shift symmetry acting on the
quiver.  The $B$ generator may be thought of as a clock generator, which
acts by rephasing the links of the quiver.  This combination of clock and
shift then generates the finite Heisenberg group as indicated above, where
$C$ is essentially a uniform rephasing of all the links.  These
transformations satisfy three important properties: $(i)$ they leave the
superpotential invariant; $(ii)$ they satisfy anomaly cancelation constraints
for all $SU(N)$ gauge groups; and $(iii)$ the above group relations are true
up to elements in the center of the gauge group $SU(N)^p$, that is, up to
gauge transformations.

The generators $A$, $B$ and $C$ can be interpreted, in the dual string theory,
as operators counting the number of wrapped F-strings, D-strings and D3 branes
respectively. Thus, the above Heisenberg group implies that the charges of
D branes in the presence of Ramand-Ramond flux do not commute.
In fact, an important motivation for our study is provided by recent investigations along these lines due to Belov, Moore and others
\cite{kitp,dima}.

An extension of this structure was considered in \cite{Burrington:2006aw} were nonconformal quiver gauge theories were considered.  In this case, the
Heisenberg group gets centrally extended as a result of having gauge groups
with different ranks. In other words, the addition of fractional branes to the
background induces a central extension of the Heisenberg group.

We believe the study of discrete symmetries of quiver gauge theories is interesting in its own right. As field theories, a natural question that arises
in this context is the interplay between the existence of a finite Heisenberg
group and Seiberg duality. Seiberg duality is an equivalence between two gauge
theories \cite{sd}, and has been extensively studied in the context of quiver
gauge theories \cite{seibergquiver,td1,trees,td3}.

In this paper we study the interplay between the existence of a finite Heisenberg group acting on orbifold quiver gauge theories and Seiberg duality.
The generator $A$, realized as a shift symmetry, acts manifestly only on
the symmetric phase of the theory.  After Seiberg duality, most quivers lose
this manifest shift symmetry associated with $A$.  However, we demonstrate that
this symmetry is naturally restored in the space of all Seiberg dual quivers.

Our main result is as follows: {\it For a quiver gauge theory admitting the action of a finite Heisenberg group
${\rm Heis}(\mathbb{Z}_q\times \mathbb{Z}_q)$ as above (\ref{heis}),
there exists a similar finite Heisenberg group acting on the space of
Seiberg dual theories to the quiver. Moreover, the action of the shift generator in the Heisenberg group maps fields among different Seiberg phases.}

The paper is organized as follows. In section \ref{generalities} we review some
known aspects of quiver gauge theories including the algorithm for performing
Seiberg duality in quiver gauge theories and some of its properties. Also in
section \ref{generalities} we provide a constructive proof of the existence of
a Heisenberg group acting on the spaces of Seiberg dual quivers. Section 
\ref{exampSect} contains several explicit examples of our construction; we
consider $\mathbb{C}^3/\mathbb{Z}_3$, $Y^{4,2}$ and $Y^{6,3}$ quivers.
We then conclude in Section \ref{conclusions}.

%%%%%%%%%%%%%%%%%%%%%%%%%%%%%%%%%%%%%%%%%%%%%%%%%%%%%%%%%%%%%%%%%%%%%%%%%%
\section{Seiberg duality and discrete symmetries}\label{generalities}
%%%%%%%%%%%%%%%%%%%%%%%%%%%%%%%%%%%%%%%%%%%%%%%%%%%%%%%%%%%%%%%%%%%%%%%%%%
In this section we begin with a discussion of generalities of Seiberg duality
acting in quiver gauge theories.  We then examine the implication of
Seiberg duality on the $B$ and $C$ generators, which act by rephasings on
the links.  As discussed in \cite{grw,Burrington:2006uu,Burrington:2006aw},
these abelian generators may be constructed by assigning discrete $U(1)$
charges to each node in the quiver consistent with anomaly cancellation.
The anomaly cancelation requirement has a natural description in terms of
the adjacency matrix, and we demonstrate how this carries over into the
Seiberg dual phases as well.  As a result, the $B$ and $C$ generators
have a natural extension when acting on the space of Seiberg duals.  By
including the $A$ generator (now acting on the space of Seiberg duals), we
are then able to prove that the space of Seiberg dual quivers naturally
inherits the action of the finite Heisenberg group of the original
(symmetric phase) quiver.

%%%%%%%%%%%%%%%%%%%%%%%%%%%%%%%%%%%%%%%%%%%%%%%%%%%%%%%%%%%%%%%%%%%%%%%%%%
\subsection{Seiberg duality on quivers}
%%%%%%%%%%%%%%%%%%%%%%%%%%%%%%%%%%%%%%%%%%%%%%%%%%%%%%%%%%%%%%%%%%%%%%%%%%
Seiberg duality has been discussed extensively in the context of quiver gauge theories. In this subsection we review its implementation and discuss some of its implications for quiver gauge theories.
We consider only quiver gauge theories with gauge group $SU(N_i)$. In this
case, a quiver gauge theory is completely defined by giving the rank of the gauge groups, the matter content and the superpotential. The matter content
is conveniently encoded in the (antisymmetric) adjacency matrix denoted by
$\mathbf{a}_{M N}$.

A given Seiberg duality acts on a single node of the quiver.  Thus, to
state the rules of Seiberg duality, it is convenient to split the indices
(labels for the nodes) according to: $0$ to indicate the specific node under
investigation; $i,j,\cdots$ to indicate neighboring nodes that have arrows
pointing {\it from} the $0^{\rm th}$ {\it to} the $i^{\rm th}$ node;
$\tilde{i},\tilde{j},\cdots$ to denote neighboring nodes that have
arrows pointing {\it from} the $j^{\rm th}$ {\it to} the $0^{\rm th}$ node;
and $a,b,\cdots$ to denote the remaining nodes, which
we emphasize are not directly connected to the $0^{\rm th}$ node.
Seiberg duality of the $0^{\rm th}$ node is captured by the
transformation
\bea
\hba_{0 M}&=&- \bba_{0 M}, \nn \\
\hba_{i j}&=&\bba_{i j}, \nn \\
\hba_{\tilde{i} \tilde{j}}&=& \bba_{\tilde{i} \tilde{j}}, \nn \\
\hba_{aM}&=& \bba_{aM}, \nn \\
\hba_{i \tilde{j}}&=& \bba_{i \tilde{j}} - \bba_{0 i} \bba_{\tilde{j} 0}, \nn \\
\hba_{\tilde{j} i}&=& \bba_{\tilde{j} i} +\bba_{0 i} \bba_{\tilde{j} 0}, \eea
where the caret denotes the new quantities after Seiberg
duality.  All other components follow from the antisymmetry of
$\bba$ and $\hba$.  If we further make the assumption that any two nodes are
only connected by edges with the same directionality, the adjacency
matrix completely determines the field content and charges of the
new theory. In addition, the rank of the $0^{\rm th}$ gauge group
is now changed to be
\be \hat{N}_0=\sum_i(\bba_{0 i} N_i) - N_0
=\sum_{\tilde{j}}(\bba_{\tilde{j} 0} N_{\tilde{j}}) - N_0.
\label{rankchange} \ee
Here we use the notation $N_M$ to denote the rank of the gauge
group at the $M^{\rm th}$ node. Note that the second equality
follows as a result of the original anomaly cancelation condition,
and that the sum term above is easily understood as the effective
number of flavors.

%%%%%%%%%%%%%%%%%%%%%%%%%%%%%%%%%%%%%%%%%%%%%%%%%
\subsection{Discrete transformations and Seiberg Duality}
%%%%%%%%%%%%%%%%%%%%%%%%%%%%%%%%%%%%%%%%%%%%%%%%%%%%%%%%

The $B$ and $C$ operators both act by rephasings of the links of the quiver.
A convenient manner to construct these operators is to assign discrete $U(1)$
charges to the nodes, say charge $n_K$ for the $K^{\rm th}$ node.  These
charges are not arbitrary, but must satisfy the constraint of anomaly
cancellation.  (The superpotential constraint is automatically satisfied,
since it corresponds to closed loops along the links.)  Here we show that this
anomaly cancellation constraint has a natural formulation in terms of the
adjacency matrix $\bba$.  However, once we do this, it then becomes clear
how to extend $B$ and $C$ to act on the Seiberg dual quivers.  The result
essentially follows by replacing the adjacency matrix $\bba$ with the
corresponding dual one $\hba$.  This results in a map of the charges $n_K$
to $\hat n_K$ in a similar fashion as (\ref{rankchange}).

Before turning to the $B$ and $C$ rephasing operators, however, we first
display the general feature of cubic anomaly cancelation for $SU(N)^3$
and explicitly demonstrate that it is still met in the Seiberg dual phase.
This calculation will then be extended to the discrete rephasing case in
a straightforward manner.  To begin, we note that the anomaly cancelation
condition in the original Seiberg phase is given by
\bea \bba_{M 0} {N}_0 + \sum_{i}(\bba_{M i} N_i) +
\sum_{\tilde{j}}(\bba_{M \tilde{j}} N_{\tilde{j}}) \sum_{a}(\bba_{M
a} N_{a})=0.
\eea
Given this condition, we wish to show that
\bea \hba_{M 0} \hat{N}_0 + \sum_{i}(\hba_{M i} \hat{N}_i) +
\sum_{\tilde{j}}(\hba_{M \tilde{j}} \hat{N}_{\tilde{j}})
\sum_{a}(\hba_{M a} \hat{N}_{a})=0, \label{toshow} \eea
which is the equivalent statement in the Seiberg dual phase.
The last summation factor in either of the above is unaffected by Seiberg
duality of the $0^{\rm th}$ node.  Hence we will denote this term simply as
$\sum_{a}(\hba_{Ma} N_{a})=\sum_{a}(\bba_{M a} N_{a})=\Sigma$, because
$\hba_{Ma}=\bba_{M a}$. In fact, as a result of this, the anomaly condition
is trivially met by all nodes except for the $0^{\rm th}$ and
those connected to it. Therefore, we will only consider the $M=k$
and $M=\tilde{k}$ terms.  For $M=k$, the left hand side of
(\ref{toshow}) becomes
\bea && \kern-4em -\bba_{k 0} \left(\sum_i(\bba_{0 i} N_i) -
N_0\right) + \sum_{i}(\bba_{k i} N_i) +
\sum_{\tilde{j}}\left((\bba_{k \tilde{j}}-\bba_{0k}\bba_{\tilde{j}0})
N_{\tilde{j}}\right) +
\Sigma \nn \\
&=& \bba_{k 0} N_0 + \sum_{i}(\bba_{k i} N_i)  +
\sum_{\tilde{j}}(\bba_{k \tilde{j}} N_{\tilde{j}}) + \Sigma
-\bba_{k 0} \sum_i(\bba_{0 i} N_i)-\sum_{\tilde{j}}(\bba_{0k}\bba_{\tilde{j}0} N_{\tilde{j}})   \nn \\
&=& -\left(\bba_{k 0} \sum_i(\bba_{0 i} N_i)+\sum_{\tilde{j}}(\bba_{k0
}\bba_{0\tilde{j}} N_{\tilde{j}})\right)
\label{zeroEcalc} \\
&=&0,  \nn \eea
where the terms are $0$ as a result of the $k^{\rm th}$ and
$0^{\rm th}$ component of the original anomaly condition.  This
condition is likewise met by the $M=0$ node because the only terms
appearing simply flip sign.  The $\tilde{j}$ nodes follow in
exactly the same manner as the above calculation because of
relation (\ref{rankchange}).  This shows that anomaly cancellation
holds in the Seiberg dual phase, so long as it holds in the original phase.

We now turn to the discrete $U(1)$ rephasings used to construct the $B$ and
$C$ operators.  Because these are abelian rephasings, we need to consider
mixed anomalies, where the anomaly comes from the $j^\mu
SU(N)^2$ triangle diagram where $j^{\mu}$ is the conserved current
of the $U(1)$ that the rephasing is associated with. To
accomplish the rephasing, we associate the phase $\omega_K$ and
charge $n_K$ with the $K^{\rm th}$ node.  A field represented by
an arrow going from the $K^{\rm th}$ node to the $L^{\rm th}$ node
gets rephased by $\omega_K^{n_K}\omega_{L}^{-n_L}$, and we mean
this component by component in the superfield.

We now consider an instanton number 1 at the $M^{\rm th}$ node, and so
all fields represented by an arrow with either end on that node have a
fermion zero mode (again, counting the other end of the arrow as
an effective flavor symmetry). We follow the previous work
\cite{Burrington:2006uu,Burrington:2006aw} and find that the
general expression for the anomaly with this instanton number is
\bea (\omega_0^{n_0} \omega_M^{-n_M})^{N_0 \bba_{0M}}
\prod_j(\omega_j^{n_j} \omega_M^{-n_M})^{N_j \bba_{jM}}
\prod_{\tilde{j}}(\omega_{\tilde{j}}^{n_{\tilde{j}}}
\omega_M^{-n_M})^{N_{\tilde{j}} \bba_{{\tilde{j}}M}}
\prod_{a}(\omega_{a}^{n_{a}} \omega_M^{-n_M})^{N_{a} \bba_{a M}}=1,
\eea
where the equality is to be read as a requirement of the set of
numbers $n_K$ and phases $\omega_K$. One can easily see that the
factors of $\omega_M$ cancel in this expression because the $N_M$
are a zero eigenvector of the adjacency matrix $\bba$, and follows
because the effective number of arrows in and out are the same.  The
expression simplifies to
\bea (\omega_0^{n_0 N_0 \bba_{0M}}) \prod_j(\omega_j^{n_j N_j
\bba_{jM}}) \prod_{\tilde{j}}(\omega_{\tilde{j}}^{n_{\tilde{j}}
N_{\tilde{j}} \bba_{{\tilde{j}}M}}) \prod_{a}(\omega_{a}^{n_{a}
N_{a} \bba_{a M}})=1. \eea
The above expression is general, and valid for any assignments of
numbers and phases. As a further simplification, we wish
to express all phases in terms of a single phase, $\omega$.  We
note also that the phase $\omega_K$ and the number $n_K$
overparameterize exactly how each field is charged.  We take that
the phases are all $e^{i \phi_K}$ where $\phi_K\in (0\cdots
2\pi]$, so that $1$ is parameterized only by $e^{2\pi i}$. One can
vary $\omega_K$ and $n_K$ while leaving $\omega_K^{n_K}$ fixed.
Therefore, we find it useful to tune all of the $\omega_K$ such
that
\be \omega_K=\omega^{1/N_K}, \label{genroot} \ee
for some fixed $\omega$ that is independent of $K$, leaving all
parametrization of the phases represented by the charges $n_K$.
Again, if $\omega=1$ we parameterize this as $\omega=e^{2\pi i}$
such that the roots above defining $\omega_K$ are non trivial (we
will return to this case in a moment). This greatly simplifies our
expression, and we find
\bea &&\kern-2em (\omega^{n_0 \bba_{0M}}) \prod_j(\omega^{n_j
\bba_{jM}}) \prod_{\tilde{j}}(\omega^{n_{\tilde{j}}
\bba_{{\tilde{j}}M}})
\prod_{a}(\omega^{n_{a} \bba_{a M}})=1     \nn \\
&=& \omega^{n_0 \bba_{0M} + \sum_j({n_j \bba_{jM}}) +
\sum_{\tilde{j}}({n_{\tilde{j}} \bba_{{\tilde{j}}M}})+
\sum_{a}({n_{a} \bba_{a M}})}=1. \eea
This, then, is the {\it general} expression for the anomaly when
the $M^{\rm th}$ node has an instanton number $1$.  We will
abbreviate the exponential above as $\overleftarrow{n}\cdot \bba$
for obvious reasons.

We therefore require that this condition is met by the numbers
$n_K$ for all values of $M$ above. One may worry that this does
not capture all possible instanton numbers.  However, because the
Pontryagin number for composite connections is additive, the basis
of taking instanton number $1$ for each node spans the space of
all possible instanton numbers, and satisfying the anomaly is most
restrictive for these individual basis vectors.  One may see that
the Pontryagin number is additive because the two triangle
diagrams $j^\mu SU_1(N_1)^2$ and $j^\mu SU_2(N_2)^2$ exist (for a
field charged under both of these gauge groups), but no cross term
exists.  Therefore, a fermionic field charged under both
$SU_1(N_1)$ and $SU_2(N_2)$ will have $J_1\times N_2+J_2\times
N_1$ fermion zero modes, where $J_i$ are the instanton numbers of
the $SU_i(N_i)$ gauge groups.  The extra factors of $N_i$ show up
as a result of the trace over the gauge indices not ``coupled to''
in the triangle diagram.

A few words are now in order to discuss
possible solutions to the above equations. If one picks the
$\overleftarrow{n}$ to be a zero eigenvector of the adjacency
matrix, there is no more condition on $\omega$ and hence $\omega$
is arbitrary. This is a full $U(1)$ symmetry that is
non-anomalous, and explains why the $D$ and $E$ operations of
\cite{Burrington:2006uu,Burrington:2006aw} were found to be simply
$\zet_p$ subgroups of these continuous $U(1)$ factors.  This also
suggests how to take the rephasings of this kind through Seiberg
duality: one reassigns $\hat{\omega}_0=\omega_0^{N_0/\sum_i
\bba_{0 i}N_i -N_0}$, and then reassign $\hat{n_0}= \sum_i \bba_{0
i}n_i -n_0$ (all other $n_M$ remain the same) and one again gets a
$U(1)$ in the new Seiberg phase.  One should note that the nodal
charge of the fields remains unchanged in this case because the
reassignment of the value of $n_0$ exactly cancels the change in
the value of $\omega_0$ so that $\omega_0^{n_0}=\hat{\omega}_{0}^{\hat{n}_0}$.

Let us now consider when the $n_M$ are integers, a case that
matches the $B$ and $C$ operations of
\cite{Burrington:2006uu,Burrington:2006aw}.  One can see that if
one takes the $n_M$ such that
\be GCD(\{(\overleftarrow{n}\cdot \bba)_M\})=\lambda, \ee
then we may simply require that
\be \omega^{\lambda}=1, \ee
and we again have a symmetry.  This time, however, the symmetry is
not continuous, but is a $\lambda^{\rm th}$ root of the center of
the gauge group.  If we label this rephasing as $Q$, we have that
$Q^{\lambda}=1$ up to the center of the gauge group.  Certainly a
class of such vectors is possible if $\bba$ has any (non zero)
integer eigenvalues.  Also note that the case $\lambda=1$
corresponds to the $\omega=1$ case previously mentioned.  As
stated, we parameterize this by $\omega=e^{2\pi i}$ such that the
roots (\ref{genroot}) are non trivial.  This is easily seen to be
a rephasing using the center of the gauge group, because the additional
root of (\ref{genroot}) makes this an $N_i$ root of for a node with rank $N_i$.  These
rephasings are therefore gauge equivalent to 1.
This also means that the integers
$n_K$ in this case are only understood modulo
$\lambda$, as one may shift the center of each gauge group independently.

Let us show how such symmetries map through Seiberg duality.
First, in the original Seiberg
phase, we assume that there is a solution to
\be \bba \cdot \overrightarrow{n}\equiv 0 \; (\mathrm{mod}\; \lambda). \ee
We therefore wish to find the new vector
$\hat{\overrightarrow{n}}$ after Seiberg duality that satisfies
\be \hba \cdot \hat{\overrightarrow{n}}\equiv 0 \; (\mathrm{mod}\;
\lambda). \ee
We again suppose that we only wish to change $\hat{n}_0\neq n_0$,
leaving all other integers alone $\hat{n}_M=n_M,M\neq0$. We now
note that again the $M=a$ components of the above equation are
automatically satisfied: only the anomalies away from the node
being dualized are affected.  We now make an educated guess as how
one transforms $n_0$.  We guess that
\bea
\hat{n}_0&=&\sum_i(\bba_{0i}n_i)-n_0     \nn \\
&=&\sum_i(\bba_{0i}n_i)+\sum_a(\bba_{0a}n_a)-n_0  \\
&\equiv&\sum_{\tilde{i}}(\bba_{\tilde{j} 0}n_{\tilde{j}})-n_0 \nn \quad
(\mathrm{mod} \; \lambda), \\
\eea
where we have used $\bba_{0a}=0$ and the original anomaly
cancelation conditions.  The calculation now goes through {\it
exactly} as it did for the zero eigenvectors (\ref{zeroEcalc}):
all $=$ signs are simply replaced with $\equiv_{(\mathrm{mod} \;
\lambda)}$.  Actually, one may expect this structure from the
discussions of \cite{Csaki:1998vj}.

We now make one final comment.  Seiberg duality of a given node
is it's own inverse, and in fact the rephasings discussed above
transform into themselves up to the center of the gauge group.
This is noted simply by the fact that after two Seiberg duals
of one node, the phase associated with the node has gone from
\be
\omega_0\rightarrow \omega_0^{\frac{N_0}{\sum_{i}\bba_{0i}N_i-N_0}}\rightarrow
\omega_0^{\frac{N_0}{\sum_{i}\bba_{0i}N_i-N_0}\frac{\sum_{\tilde{j}}\bba_{\tilde{j}0}N_{\tilde{j}}-N_0}{N_0}}=\omega_0,
\ee
where in the second Seiberg duality we used that what we were calling ``in'' arrows are now
called ``out'' arrows, hence the switch in the sum from indices with no
tilde to ones with tilde.  This is of course the same because the effective number
of in and out arrows are the same, but this will be more important
for the mapping of the numbers $n_M$.  For this, we note that
the number $n_0$ gets mapped as
\bea
n_0 && \rightarrow
\sum_{i}\bba_{0i}n_i-n_0  \nn \\
&&\rightarrow \sum_{\tilde{j}}a_{\tilde{j}0}N_{\tilde{j}}-(\sum_{i}\bba_{0i}n_i-n_0) \nn \\
&&=n_0+\sum_{\tilde{j}}a_{\tilde{j}0}N_{\tilde{j}}-\sum_{i}\bba_{0i}n_i  \nn \\
&&\equiv n_0\;\;(\mathrm{mod} \; \lambda).
\eea
In the last line, we have again used the fact that we are modding out
by the center of the gauge group, which corresponds to taking the
number $n_M$ as only being defined $mod\; \lambda$.

\subsection{General proof of the existence of the Heisenberg Group}

Given the natural mapping of discrete $U(1)$ charges under Seiberg
duality found above, we now combine this with the
results of \cite{Burrington:2006uu,Burrington:2006aw}
to show that there is generically an action of the Heisenberg
group on the space of Seiberg dual quivers.  Here, we are assuming that
there is a symmetric Seiberg phase
with a natural shift symmetry, which we will call $A$ (see
section \ref{exampSect} for some examples).  A large class
of these was examined in \cite{Burrington:2006uu,Burrington:2006aw}
and a the action of a finite Heisenberg group was found.
Assuming that there is a symmetric phase, and an action
of a Heisenberg group on this phase, we will show the
existence of an action of a Heisenberg group on the
space of all Seiberg phases.

First, note that
if there is an $A$ symmetry of the symmetric phase,
this descends to the entire tree of possible Seiberg
dual quivers, however mapping from one phase to another.
This space was discussed in \cite{trees} where it was given the name of duality tree;  each point
represents a quiver. In particular the
three-nod tree was shown to be related to Markov's equation.
We will be careful to call the {\it points}
on the duality tree {\it points}, which represent entire
quivers, to distinguish them from the {\it nodes} of the quivers
themselves which represent gauge groups.
The action on the duality tree can be seen easily.  Take that we have a
symmetric phase with $x\times y$ nodes, and there
is a manifest $\zet_x$ symmetry that permutes
the $M^{\rm th}$ node to the $(M+y)^{\rm th}$ node.
The entire tree
is defined by taking an arbitrary number of Sieberg
dualities on any number of the nodes in any given order.
We therefore label the point on the Markov tree that is $\ell$
Seiberg dualities away from the symmetric phase by the Seiberg dualities it takes to get there: $(s_1,s_2,s_3,\cdots s_{\ell})$.  Here, the integers $s_i$
label which node of the quiver diagram is to be Seiberg dualized,
i.e. $s_i\in {1\cdots x\times y}$.
There is an
analogous quiver given by $(s_1+y,s_2+x,s_3+y,\cdots s_{\ell}+y)\equiv \overrightarrow{s}+\overrightarrow{y}$
with the same matter content, same gauge groups and couplings,
simply with the labels changed.  This, therefore, defines the
action of the $A$ operator on this quiver: it maps the two
differently labeled field theoretic degrees of freedom into
each other.  So, if one wants to go
from one branch to the other, one must apply a set of inverse Seiberg dualities
to get back to the symmetric phase, apply the $A$ operation (as many times
as needed) to reorder the fields, and then apply the same set of Seiberg
dualities.  However, because the ordering of the fields have been switched,
the Seiberg dualities are in fact being performed along a different branch of the
duality tree.  We emphasize here that
this is much the same as the operation $B$ where the rephasings
are really on the order one gives the fields in, not their subscripts :$A$
may have already rearranged them, or may not have. $B$'s matrix representation
does not depend on this!  So, when we say that ``$B$ is rephasing the field
$U_1$ as $u_1\times U_1$'', we really mean that it rephases this way only
when the fields are given in the canonical order $(U_1,U_2,\cdots)$; it really
rephases the {\it first} field in the list.  $A$ of course changes exactly which
field this is.  We think of
the Seiberg dualities in the same way.  We give the gauge groups and fields
a canonical ordering which $A$ shuffles.  The $S_{\vect{s}}$ simply Seiberg
dualize according to $\vect{s}$ in the order that the gauge groups are listed,
not what their labels are, and so may or may not move out along a different branch,
depending on whether an $A$ is present or not.

We will make these comments more precise here.
We have labeled the series
of Seiberg dualities with a vector $\overrightarrow{s}$, and
we will call the series of Seiberg dualities $S_{\overrightarrow{s}}$.
The inverse is of course given by the same vector, simply with its
entries reversed $S_{\overrightarrow{s}}^{-1}= S_{{\mbox{reverse order}}(\overrightarrow{s})}^{-1}$.
$A$ is simply defined using the symmetric phase
\bea
\hat{A}_{\overrightarrow{s}', \overrightarrow{s}} &=&\delta_{\vect{s}',\vect{s}}S_{\overrightarrow{s}}A S_{\overrightarrow{s}}^{-1}
\eea
with all other entries for $\overrightarrow{s}$ and $\overrightarrow{s}'$ equal
zero.  One must be careful here to note that something new has actually happened.  The
$S^{-1}$ that maps back to the symmetric phase and the $S$ that maps one out are (as matrices)
identical.  However, because the fields on which they act have been shifted by $A$ one is
actually going out on a different branch of the duality tree.  It is equivalent
to going out on the one shifted by the $\vect{y}$ vector.  The
rephasing operations $B$ and $C$ may be similarly defined as:
\bea
\hat{B}_{\overrightarrow{s}, \overrightarrow{s}'}\equiv S_{\overrightarrow{s}}B S_{\overrightarrow{s}}^{-1}
\delta_{\overrightarrow{s}, \overrightarrow{s}'} \nn \\
\hat{C}_{\overrightarrow{s}, \overrightarrow{s}'}\equiv S_{\overrightarrow{s}}C S_{\overrightarrow{s}}^{-1}
\delta_{\overrightarrow{s}, \overrightarrow{s}'},
\eea
where in the last section we have shown how one maps these rephasings through a
Seiberg duality explicitly
(they are diagonal on the space of Seiberg phases because the $B$ operation
simply rephases the fields, and does not mix different fields).  Hence one should
think of $\hat{B}$ as working on
all points of the duality tree simultaneously, and one thinks of $\hat{A}$ as mapping
the entire tree into itself.

The above operators satisfy the Heisenberg group structure
on the duality tree.  Let us consider a general quiver in the tree
$Q_{\overrightarrow{s}_1}$ and the
action of $\hat{A},\hat{B}$ and $\hat{C}$ on this quiver
\bea
&&\kern-2em \hat{A}_{\vect{s},\vect{s}'}\hat{B}_{\vect{s}', \vect{s}_1}Q_{\vect{s}_1} \nn \\
&&=\delta_{(\vect{s},{\vect{s}_1})}S_{\vect{s}}A B S_{\vect{s}_1}^{-1} Q_{\vect{s}_1} \nn \\
&&=\delta_{(\vect{s},{\vect{s}_1})}S_{\vect{s}}B A C S_{\vect{s}_1}^{-1} Q_{\vect{s}_1} \\
&&=\hat{B}_{\vect{s},\vect{s}'} \hat{A}_{\vect{s}',\vect{s}''} \hat{C}_{\vect{s}'',\vect{s_1}} Q_{\vect{s}_1}  \nn
\eea
where we make a special note that there is only one non zero entry for each $\vect{s}$
and $\vect{s}'$ such that that the only implicit summation is over those $\vect{s}$ in the
$\delta$ symbols: the indices in the Seiberg duals are {\it fixed}.  All of them are in fact
exactly equal to $\vect{s}_1$ because they each contain a Kronecker $\delta$.  It is again
the action of $A$ which is non trivial and mixes the fields, so exactly which gauge group
the entries of $\vect{s}$ are referring to depend only on the order in which the gauge groups are
listed, which is changed by the presence of an $A$.
In fact all of the relations found in the symmetric phase
descend to the entire tree, and so we find
\bea
\hat{A}\hat{B}=\hat{B}\hat{A}\hat{C},\quad \hat{C}\;\;{\mbox{commutes with everything}},
\quad \hat{A}^{x}=\hat{B}^x=\hat{C}^x=1.
\eea
This, then, is an action on the whole space of possible Seiberg dual theories, and the equals sign
are read only up to the center of the gauge group.  The ranks and possible gauge couplings are
different in each Seiberg phase, and so we take that the gauge redundancies are modded out in each
phase individually.  There may be some understanding of this as some duality among the relevant
Faddeev-Popov ghosts, however we satisfy ourselves here by simply noting that the extra factors
are purely gauge.

%%%%%%%%%%%%%%%%%%%%%%%%%%%%%%%%%%%%%%%%%%%%%%%%%%%%%%%%%%
\section{Finite Heisenberg groups acting on the space of Seiberg dual quivers}
\label{exampSect}
%%%%%%%%%%%%%%%%%%%%%%%%%%%%%%%%%%%%%%%%%%%%%%%%%%%%%%%%%

\subsection{$\mathbb{C}^3/\mathbb{Z}_3$}
\begin{figure}[h]
\begin{center}
\epsfig{file=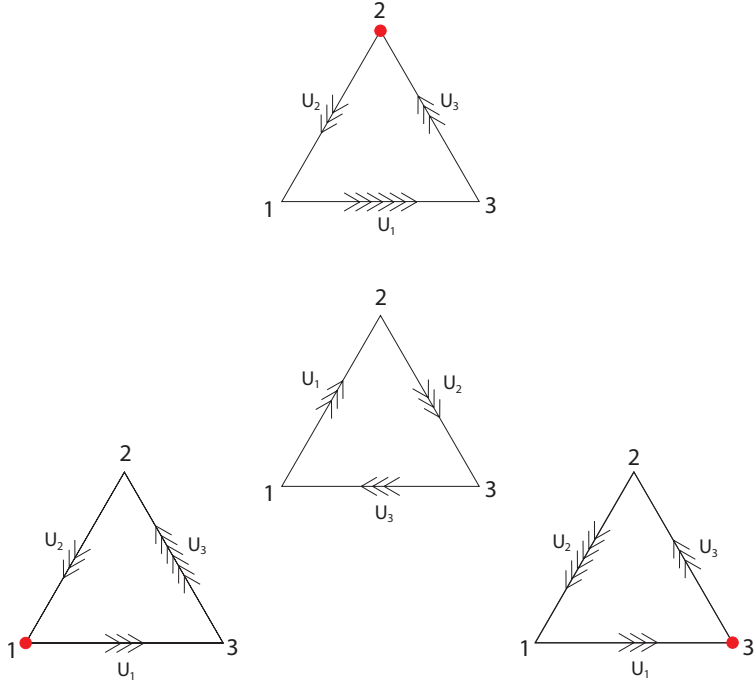,width=10 cm} \caption{$\mathbb{C}_3/\mathbb{Z}_3$ quivers. Each Seiberg dual does not possess a $\mathbb{Z}_3$
symmetry. However, the space of Seiberg duals has a manifest shift symmetry generated by $\mathbb{Z}_3$.}
\label{figC3}
\end{center}
\end{figure}

Let us first consider a simple example discussed originally in \cite{grw}. We take the
gauge theory dual to the Maldacena limit of string theory on $\mathbb{C}^3/\mathbb{Z}_3$ where the
orbifold action is given by $(z_1,z_2,z_3) \to (\xi z_1,\xi z_2,\xi^{-2} z_3)$ where $\xi$ is a cubic root of unity.
The quiver
diagram is represented in the center of the figure (\ref{figC3}).
Let the rank of each of the gauge group be $N_c$. Performing
Seiberg duality at the node doubles its rank and the node is
represented by a red circle. Three different quivers can be
obtained by performing duality at the 3 different nodes (See
Figure \ref{figC3}). We will show in this section that there
exists a
Heisenberg group acting on the three  Seiberg dual quivers.\\
  We label the dual quivers on the top , left and right as
T, L and R respectively. The A transformation permutes the nodes
between different Seiberg dual quivers. Let it act as \bea
1L\rar 2T\rar 3R, \nn\\
2L \rar 3T \rar 1R,\nn\\
3L \rar 1T \rar 2R, \eea
where the number in the front denotes the
node corresponding to the quiver represented by the alphabet next
to it.
B and C transformations are phase
transformations of chiral fields. Let us denote the phase of field
$U_i$ by $u_i$. These phases respect invariance of the superpotential  and
anomaly cancelation. These relations for the phases in the left
quiver are
\bea
u_1u_2u_3&=&1,\\
(u_1^3u_2^3)^N&=&1,\nn\\
(u_2^{3.2}u_3^6)^N&=&1,\nn\\
(u_3^6u_1^{2.3})^N&=&1.
 \eea
 They are further simplified to obtain
 \be
 u_3=(u_1u_2)^{-1}\;\;\;\;(u_1u_2)^{3N}=1\;\;\;\;u_1^{6N}=u_2^{6N}=1
 \ee
 Below we write a particular solution for the phase assignments.
 We write it in terms of $3N-$th root of unity denoted as $\omega$ i.e.
 ${\omega}^{3N}=1$. The topmost row denote the subscript corresponding to the
fields whose phases are written below in that column.
 The numbers in the rows
corresponding to B and C denote the powers to which $\omega$ is
raised. \be
\begin{array}{|c|c|ccc|}\hline &&1&2&3\\
\hline &L&0&0&0\\
B&T&0&1&-1\\
&R&1&0&-1\\
\hline &L&-1&1&0\\
C&T&0&-1&1\\
&R&1&0&-1\\
\hline
\end{array}\ee
Let us make a general remark about a technical point.
The generators of the Heisenberg group satisfy
\bea AB&=&BAC,\nn\\
AC&=&CA,\nn\\
BC&=&CB,\nn\\
\label{group}
A^q=B^q=C^q&=&1,
\eea
up to an element in center of
the gauge group. These can be rewritten as
\bea
C^{-1}A^{-1}B^{-1}AB&=&Z_1^c\nn\\
A^{-1}C^{-1}AC&=&Z^c_2\nn\\
B^{-1}C^{-1}BC=1\nn\\
A^q=1\nn\\ B^q=Z^c_3\nn\\
\label{heisen} C^q=Z^c_4 \eea
 where
$Z^c_i$s are elements in center of the gauge group.

Thus, after explicitly constructing the elements $A,B$ and $C$, we need to present the central
elements involved in the construction of the group.
The particular elements in the center of the gauge group needed
for the Heisenberg group can be written in terms of a set of 3
integers $(a_1, a_2, a_3)$. These three elements will stand for a
rephasing at the nodes (1,2,3) by ${\omega}^{3a_1/2}, {\omega}^{3a_2},
{\omega}^{3a_3} $ respectively for the case of the left quiver
diagram. The factor of $1/2$ in the exponent of the phase
associated with node 1 is due to the fact that the rank of the
gauge group at node 1 is twice that of the others for the left
quiver diagram. The needed elements of the center of the gauge
group as denoted in equations (\ref{heisen}) can then be written
as \bl
Z^c_{4L}:&a_2=a_3=1\;\;a_1=0\nn\\
Z^c_{1T}:&a_1=a_3=-1,\;\; a_2=0\nn\\
 Z^c_{3R}:& a_1=a_2=1,\;\;a_3=0\nn\\
&Z^c_{3T}=Z^c_{1T}\;\;\;\;Z^c_{4T}=-Z^c_{1T}\;\;\;\;Z^c_{4R}=Z^c_{3R}\nn\\
&Z^c_{1L}=Z^c_{1R}=Z^c_{2L}=Z^c_{2T}=Z^c_{2R}=Z^c_{3L}=1
 \end{align*}
The center of the gauge group corresponds to the left, top and the
right quiver diagrams depending on the extra subscript $(L, T \tr{
or } R)$ respectively.

%%%%%%%%%%%%%%%%%%%%%%%%%%%%%%%%%%%%%%%%%%%%%%%%%%%%%%%%%%%%%%%%%%
\subsection{Heisenberg group in $Y^{4,2}$}
Consider the quiver diagram for $Y^{4,2}$ as shown in Fig. 2  with chiral fields $U_i, Y_i$ and $Z_i$\\
\begin{figure}[h]
\begin{center}
\epsfig{file=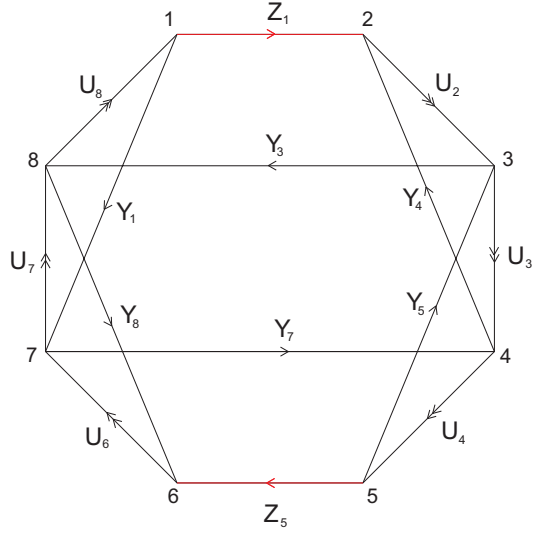,width=7 cm}
 \caption{$Y^{4,2}$ quiver diagram}
 \label{Fig. 1}
\end{center}
\end{figure}
It has a cyclic $\mathbb{Z}_2$- symmetry interchanging the nodes as
(15)(26)(37)(48). It is convenient to introduce the following notation, ${\bf
U}=[U_2,U_3,U_4,U_6,U_7,U_8]^{T}, {\bf
Y}=[Y_1,Y_3,Y_4,Y_5,Y_7,Y_8]^T$ and ${\bf Z}=[Z_1, Z_5]^T$ to
denote the sets of chiral fields. The A transformation acts on
chiral fields as \bea
A{\bf U}&=&\lf [ \ba {cc}0&I_3\\I_3 &0\ea\ri ]{\bf U}\nn\\
A{\bf Z}&=&\lf [\ba {cc}0&1\\1&0 \ea\ri ]{\bf Z}\nn\\ A{\bf
Y}&=&\lf [ \ba {cc}0&I_3\\I_3 &0\ea\ri ]{\bf Y}\eea where $I_3$ is
a 3$\times$ 3 matrix. We will try to find other symmetries which
are related to phase changes of the chiral fields. Let the phases
be represented by small case letters. The relations among the
phases arising due to invariance
of the superpotential are
\bea z_1u_2y_3u_8&=&1,\nn\\
u_2u_3y_4&=&1,\nn\\
u_3u_4y_5&=&1,\nn\\
u_4z_5u_6y_7&=&1,\nn\\
u_6u_7y_8&=&1\nn\\\label{e-y} u_7u_8y_1&=&1. \eea
These equations
determine the action on {\bf Y}.\\ ${\bf Y}\rar $
$[(u_7u_8)^{-1},(z_1u_2u_8)^{-1},(u_2u_3)^{-1},(u_3u_4)^{-1},(u_4u_6z_5)^{-1},(u_6u_7)^{-1}]^T${\bf
Y}. There are further constraints on the phases due to anomalies.
They
are
\bea (u_8^2y_1z_1)^N&=&1,\nn\\(z_1u_2^2y_4)^N&=&1,\nn\\(u_2^2y_3y_5u_3^2)^N&=&1,\nn\\
(u_3^2y_4y_7u_4^2)^N&=&1,\nn\\(u_4^2y_5z_5)^N&=&1,\nn\\
(z_5y_8u_6^2)^N&=&1,\nn\\(u_6^2y_7y_1u_7^2)^N&=&1,\nn\\(u_7^2y_8y_3u_8^2)^N&=&1.
\eea These can be simplified to obtain \bea
u_2^{2N}=u_{4}^{2N}&=&u_{6}^{2N}=u_{8}^{2N},\nn\\
u_2^N=u_6^N,&&
u_4^N=u_8^N,\nn\\
\lf (\f{u_3}{u_7}\ri )^N &=&\lf (\f{u_2}{u_8}\ri
)^N,\nn\\
\label{e-u} z_1^N=\lf (\f{u_3}{u_2}\ri )^N,&&z_5^N=\lf
(\f{u_3}{u_4}\ri )^N. \eea
Both the other generators of the
Heisenberg group, namely $B$ and $C$, are phase changes which will
act on {\bf U}, {\bf Z} and {\bf Y} diagonally. They should also
satisfy relations (\ref{e-u}) and (\ref{e-y}). A particular
solution is \bea
B: z_1=u_4=u_7=u_8=1&u_2=u_3=\omega &z_5=u_6={\omega}^{-1}\nn\\
\label{sol1}C:u_2=u_4=u_6=u_8=1&z_1=z_5=\omega&u_3=u_7={\omega}^{-1} \eea
Here, ${\omega}^{2N}=1$. Thus, we have the action of ${\rm Heis}(\mathbb{Z}_2\times \mathbb{Z}_2)$.

\subsubsection*{Center of the gauge group}
\label{Cengg}

The Heisenberg group is closed up to center of the
gauge group. It will be useful to first give a quick look at the
center. The center has 8 generators with 8 parameters ($a_i$ say),
one acting at each node. It changes the chiral field in its fundamental (anti-fundamental)
 representation by $w^{\pm a_i}$. It is convenient to work with a new set of generators $\a _i$,
 such that $\a_i=a_i-a_{i+1}\;\: (\a_{8}=a_8-a_1)$. However, there exists a relation between
them.

The center acts on the chiral fields as
\bea
\bf{U}&\rar&\tr{diag}[{\omega}^{2\a_2},{\omega}^{2\a_3},{\omega}^{2\a_4},{\omega}^{2\a_6},{\omega}^{2\a_7},{\omega}^{2\a_8}]\bf{U}\nn\\
\bf{Y}&\rar&\tr{diag}[{\omega}^{2\a_1},{\omega}^{2\a_5}]\bf{Z}\nn\\
\bf{Z}&\rar&\tr{diag}[{\omega}^{-2\a _7-2\a _8},{\omega}^{-2\a _2-2\a _8-2\a _1},{\omega}^{-2\a _2-2\a _3},{\omega}^{-2\a _3-2\a _4},{\omega}^{-2\a _4-2\a _6-2\a _5},
{\omega}^{-2\a _6-2\a _7}]\bf{Y}\nn\\
\eea
The generators $\a _i$ satisfy a relation 
\be \label{sum}
\a _1+\a _2+\a _3+\a _4+\a _5+\a _6+\a _7+\a _8=0\ee
 For the
transformations given in (\ref{sol1}), the particular elements in
the center are \bea Z_1^c:&& \a _2=\a
_3=1\;\;\;\;\a_5=\a_6=-1\;\;\;\;\a _1=\a_4=\a _7=\a_8=0\nn\\
Z_2^c:&&\a_i=0\nn\\
&&Z_3^c=Z_1^c\nn\\
Z_4^c:&&\a_1=\a _5=1\;\;\;\;\a
_3=\a_7=-1\;\;\;\;\a_2=\a_4=\a_6=\a_8=0
 \eea
%%%%%%%%%%%%%%%%%%%%%%%%%%%%%%%%%%%%%%%%%%%%%%%
\subsubsection{Seiberg duals of $Y^{4,2}$}
\begin{figure}[h]
\begin{center}
\epsfig{file=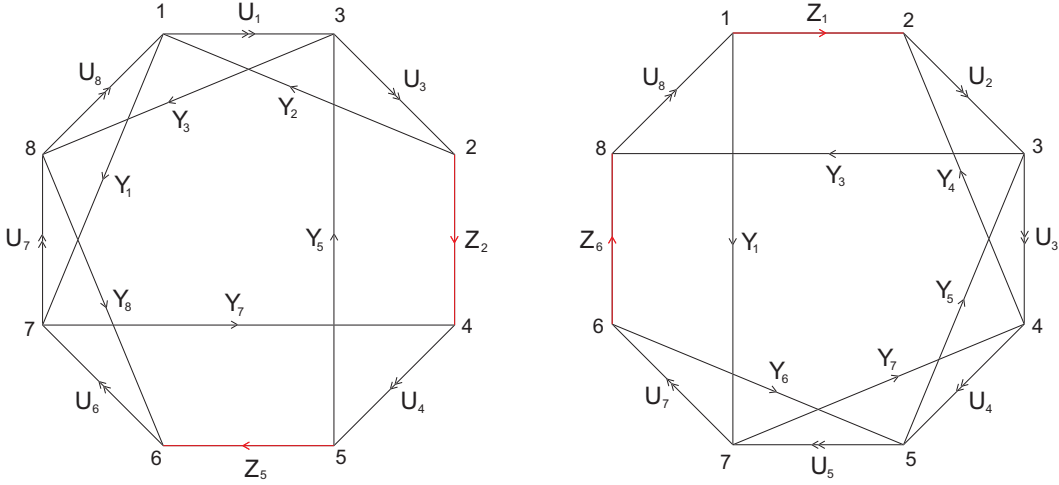,width=14 cm} \caption{Two toric phases of
$Y^{4,2}$ quiver}\label{Fig.42P}
\end{center}
\end{figure}

Toric duality in the case of $Y^{p,q}$ spaces shift the one of the singlet fields in the outside of the
quiver diagram \cite{ypqphases}. Therefore, we concentrate on two Seiberg duals phases
 of $Y^{4,2}$ in figure (\ref{Fig.42P}). They are obtained by Seiberg
dualizing at nodes 2 and at node 6. We can think of A symmetry
here as the one which takes node i of the left quiver to a node
i+4 (i-4 for $i>4$) of the right quiver on the and vice-versa. B and C
transformations are changes of phases of chiral fields, which are
constrained. For the left quiver, the superpotential conditions
are \bea y_1u_7u_8=1,&&
y_3u_8u_1=1,\nn\\
y_2u_1u_3=1,&&
y_5u_3z_2u_4=1,\nn\\
y_7u_4z_5u_6=1,&& y_8u_6u_7=1.
 \eea
The
anomaly cancelation conditions are 
\bea (u_8^2u_1^2y_1y_2)^N=1,&&
(u_3^2y_2z_2)^N=1,\nn\\
(u_1^2u_3^2y_3y_5)^N=1,&&
(u_4^2z_2y_7)^N=1,\nn\\
(u_4^2z_5y_5)^N=1,&&
(u_6^2y_8z_5)^N=1,\nn\\
(u_6^2u_7^2y_1y_7)^N=1,&& (u_7^2u_8^2y_8y_3)^N=1. \eea The former
set helps to write the phases $y_i$ in terms of the other phases.
The second set can be reduced to obtain
\bea u_1^{2N}=u_7^{2N}
\;\;u_4^{N}=u_8^N=\lf (\f{u_7}{z_2}\ri )^N,\nn\\
u_3^N=u_6^N=\lf(\f{u_1}{z_2} \ri )^N,\;\;\;\;
 z_5^N=\lf
(\f{u_1z_2}{u_7}\ri )^N. \eea
A particular solution is to consider
the phases \be
u_1=u_7=1\;\;u_3=u_4=u_6=\omega\;\;z_2=z_5=u_8={\omega}^{-1}.
 \ee
 Next , we construct the phases for the transformations B and C acting on all the chiral fields in the two quivers.
 We present them in the table below, where the topmost row are the subscripts of $u_i$ or $z_i$.
 The numbers in other rows are the powers to which $\omega$ is raised for the given transformation indicated on
  the left. The letters L and R in the second column refer to left and right quivers drawn in figure
   (\ref{Fig.42P}).  We assume ${\omega}^{2N}=1$.
 \be
\begin{array}{|c|c|cccccccc|}\hline &&1&2&3&4&5&6&7&8\\
\hline
B&L&0&-1&1&1&-1&1&0&-1\\
&R&0&0&0&0&0&0&0&0\\
\hline
C&L&0&-1&1&1&-1&1&0&-1\\
&R&-1&1&0&-1&0&-1&1&1\\
\hline
\end{array}\ee
The particular elements of the center of the gauge group needed to
satisfy the Heisenberg algebra in this case can be written in
notation used in section (\ref{Cengg}). We will use notation
$Z_i^c$ to denote various elements in the center and $\a_i$ as its
generator. We will also put a subscript L or R to denote to
dintinguish the action of the center on the two different quivers.
They necessary $Z_i^c$ are \bea
&Z^c_{1L}=Z^c_{1R}=Z_{2L}^c=Z_{2R}^c=Z_{3R}^c=I\; (\tr{identity
element}),\nn\\
Z^c_{3L}:&\a_3=\a_4=\a _6=1\;\; \a _2=\a _5=\a _8=-1\;\;\a _1=\a
_7=0,\nn\\
&Z_{4L}^c=Z_{4R}^c,\nn\\
Z^c_{4R}:&\a_2=\a_7=\a _8=1\;\; \a _1=\a _4=\a _6=-1\;\;
\a _3=\a_5=0.
 \eea
%%%%%%%%%%%%%%%%%%%%%%%%%%%%%%%%%%%%%%%%%%%%%%%%%%%%%%%%
\subsection{$Y^{6,3}$ quiver with manifest shift symmetry $A$}

We look at a toric phase of $Y^{6,3}$ quiver. It possesses A
symmetry which act on nodes as (1,5,9) (2,6,10) (3,7,11) (4,8,12).
The Heisenberg group for this case has been worked out in
\cite{Burrington:2006uu}.
\begin{figure}[h]
\begin{center}
\epsfig{file=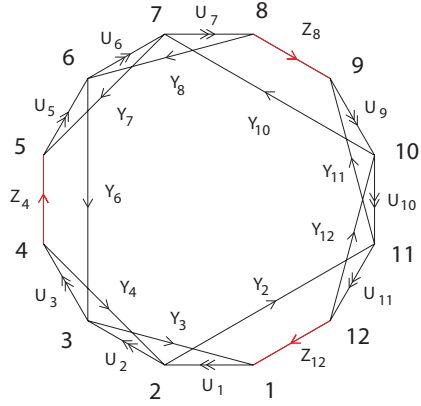,width=8 cm} \caption{Quiver gauge theory for $Y_{6,3}$ with a manifest $\mathbb{Z}_3$ shift symmetry.}
\label{Fig.4}
\end{center}
\end{figure}
The superpotential and anomaly conditions are \bea
u_1u_2y_3=1,&&u_2u_3y_4=1,\nn\\
u_3z_4u_5y_6=1,&&u_5u_6y_7=1,\nn\\
u_6u_7y_8=1,&&u_7z_8u_9y_{10}=1,\nn\\
u_9u_{10}y_{11}=1,&&u_{10}u_{11}y_{12}=1,\nn\\
u_{11}z_{12}u_1y_2=1,\\
(u_1^2y_3z_{12})^N=1,
&&(u_1^2u_2^2y_2y_4)^N=1,\nn\\
(u_2^2u_3^2y_3y_6)^N=1,&&(u_3^2z_4y_4)^N=1,\nn\\
(u_5^2z_4y_7)^N=1,&&(u_5^2u_6^2y_6y_8)^N=1,\nn\\
(u_6^2u_7^2y_7y_{10})^N=1,&&(u_7^2z_8y_8)^N=1,\nn\\
(z_8u_9^2y_{11})^N=1,&&(u_9^2u_{10}^2y_{10}y_{12})^N=1,\nn\\
(u_{10}^2u_{11}^2y_{11}y_{2})^N=1,&&(u_{11}^2z_{12}y_{12})^N=1.
 \eea
One particular solution is  \begin{align}
B:&u_1=u_6=u_7=z_8=1\nn\\&u_2=u_5=u_{11}=z_{12}=\omega\nn\\&u_{3}=z_4=u_9=u_{10}={\omega}^{-1}\nn\\
C:&u_1=u_3=z_4=u_5=u_7=z_8=u_9=u_{11}={\omega}^{-1}\nn\\&
u_2=u_6=u_{10}=\omega\nn\\&z_{12}={\omega}^{5}
\end{align}
In order to write the center of the gauge group, we will use a
similar notation as in section (\ref{Cengg}) for $Y^{4,2}$ case where
$a_i$ for $i=1...12$ denote the generator at each node.  These
generators are commutative and each of them changes the phase of
an incoming(outgoing) quiver by ${\omega}^{-3a_i}({\omega}^{+3a_i})$. We
take linear combinations of the generators as $\a_i=a_i-a_{i+1} (\alpha_{12}=a_{12}-a_1)$.
Then the elements in the center of the gauge group are
\begin{align}
Z_1^c:&\a_5=\a_{11}=1,\;\;\a_{10}=\a_{12}=-1\;\;{\tr{zero otherwise.}}\nn\\
Z_2^c:&\a_{12}=2\;\;\a_{8}=-2\;\;{\tr{zero otherwise.}}\nn\\
Z_3^c:&\a_2=\a_5=\a_{11}=\a_{12}=1\;\;\a_{3}=\a_{4}=\a_{9}=\a_{10}=-1\;\;\a_1=\a_6
 =\a_7=\a_8=0\nn\\
Z^c_4:&\a_2=\a_6=\a_{10}=1\;\;\a_{1}=\a_3=\a_4=\a_5=\a_7=\a_8=\a_9=\a_{11}=-1\;\;\a_{12}=5
\end{align}
%%%%%%%%%%%%%%%%%%%%%%%%%%%%%%%%%%%%%%%%%%%%%%%%%%%%%%%%%%%%%%%
\subsection{Seiberg duals of $Y^{6,3}$}
\begin{figure}[h]
\begin{center}
\epsfig{file=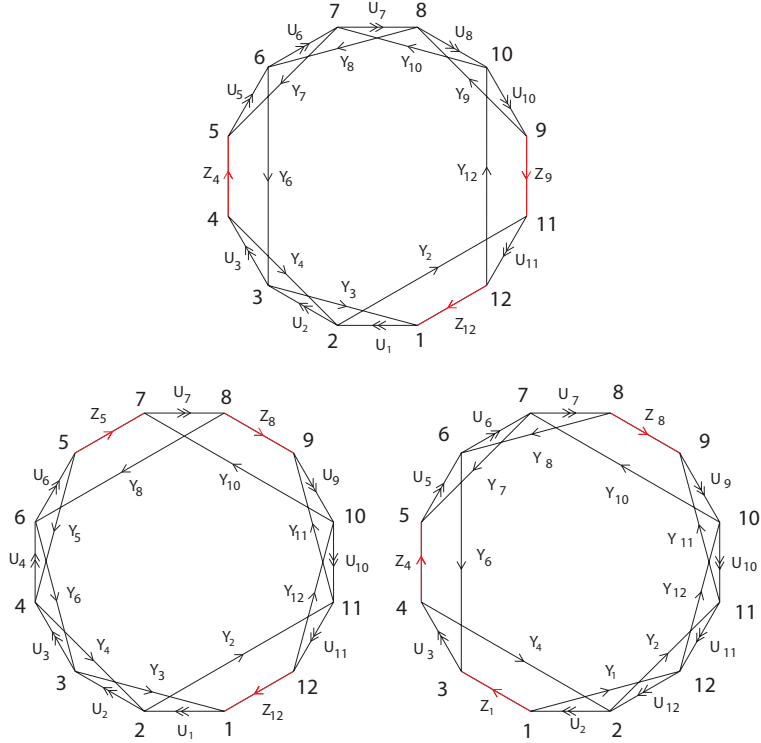,width=10 cm}  \caption{Three Seiberg phases of $Y_{6,3}$. Each phase lacks a $\mathbb{Z}_3$ shift 
symmetry but the symmetry is present when considering the space of three Seiberg phases.}\label{Fig.63P}
\end{center}
\end{figure}
Now we look at certain Seiberg duals of $Y^{6,3}$ quiver. Let us label
the figures in figure (\ref{Fig.63P}) on the top, left and right as $T$, $L$ and $R$. The A
transformation acts by taking a node i of the top(left) quiver to
a node i+4 (i-8 if $i>8$) of the right (top) quiver as well as a node i of the
right quiver to a node i+4 (i-8 if $i>8$) of the left quiver. Let us consider the quiver diagram on the left and find the 
conditions for the phases associated with the B and C transformations. Clearly, similar conditions can be written for 
the top and the right. 
\bea
y_4u_2z_1u_3=1,&&\nn\\
y_6u_3z_4u_5=1,&&y_7u_5u_6=1,\nn\\
y_8u_6u_7=1,&&y_{10}u_7z_8u_9=1,\nn\\
y_{11}u_9u_{10}=1,&&y_{12}u_{10}u_{11}=1,\nn\\
\label{sup63} y_2u_{11}u_{12}=1,&&y_1u_{12}u_2=1,\eea
\bea
(z_1u_2^2y_1)^N=1,&&(u_2^2u_{12}^2y_4y_2)^N=1,\nn\\
(z_1u_3^2y_6)^N=1,&&(u_3^2z_4y_4)^N=1,\nn\\
(z_4u_5^2y_{7})^N=1,&&(u_5^2u_6^2y_{6}y_{8})^N=1,\nn\\
(u_6^2y_{7}y_{10}u_7^2)^N=1,&&(u_7^2z_8y_{8})^N=1,\nn\\
(z_8u_9^2y_{11})^N=1,&&(y_{10}u_9^2u_{10}^2y_{12})^N=1,\nn\\\label{ano63}
(u_{10}^2u_{11}^2y_{11}y_2)^N=1,&&(u_{11}^2u_{12}^2y_{12}y_1)^N=1.
 \eea
 Here the set (\ref{sup63}) comes from superpotential invariance
 and the set (\ref{ano63}) are the anomaly constraints. The first
 set allows $y_i'$s to be solved in terms of other phases. The
 second set can be simplified to obtain
 \bea
u_3^{3N}=u_2^{3N},&&z_4^N=\lf (\f{z_1u_2}{u_3} \ri )^N,\nn\\
u^N_5=\lf (\f{u_3^2}{u_2}\ri )^N,&&u_6^N=(z_1u_3)^N,\nn\\
u_7^N=u_2^N,&&z_8^N=\lf (\f{z_1u_3}{u_2}\ri )^N,\nn\\
u_9^N=u_3^N,&&u_{10}^N=\lf (\f{z_1u_2^2}{u_3}\ri )^N,\nn\\
u_{11}^N=\lf (\f{u_2^2}{u_3}\ri )^N,&&u_{12}^N=(z_1u_2)^N.
 \eea
 For a particular solution, we will write B and C acting on $u_1$
 and $z_i$ only. The numbers in the top row of the table below denote
the subscripts of $u_i$ and $z_i$. The numbers in the rows
corresponding to B and C denote the powers to which $\omega$ is
raised. \be
\begin{array}{|c|c|cccccccccccc|}\hline &&1&2&3&4&5&6&7&8&9&10&11&12\\
\hline &R&1&0&1&0&-1&-1&0&-1&1&0&-1&1\\
B&L&0&0&0&0&0&0&0&0&0&0&0&0\\
&T&1&1&0&1&-1&0&1&-1&-1&0&-1&0\\
\hline &R&1&0&1&0&-1&-1&0&-1&1&0&-1&1\\
C&L&1&0&-1&1&1&0&1&0&-1&-1&0&-1\\
&T&-1&-1&0&-1&1&0&-1&1&1&0&1&0\\
\hline
\end{array}\ee
We will write the center of gauge group in terms of generators $\a
_i$ as defined in previous subsection. We reinterpret few of the
generators as \bea {\tr{For L quiver}}&\a_4=a_4-a_6&\a_5=a_5-a_7\;\;\a_6=a_6-a_5\nn\\
{\tr{For T quiver}}&\a_8=a_8-a_{10}&\a_9=a_9-a_{11}\;\;\a_{10}=a_{10}-a_9\nn\\
{\tr{For R
quiver}}&\a_1=a_1-a_3&\a_2=a_2-a_1\;\;\a_{12}=a_{12}-a_2\nn
 \eea
The elements in the center of the gauge group needed for the
algebra are \bea
&Z_{1R}^c=Z_{1L}^c=Z^c_{2R}=Z_{2L}^c=Z_{2T}^c=Z_{3L}^c=I\;(\tr{Identity
element}),\nn\\
Z^c_{3R}:&\a_1=\a_3=\a_9=\a_{12}=1,\nn\\&\a_5=\a_6=\a_8=\a_{11}=-1,\nn\\&\a_2=\a_4=\a_7=\a_{10}=0,\nn\\
Z^c_{4L}:&\a_5=\a_8=\a_9=\a_{11}=1,\nn\\&\a_1=\a_2=\a_4=\a_{7}=-1,\nn\\&\a_3=\a_6=\a_{10}=\a_{12}=0,\nn\\
Z^c_{1T}:&\a_3=\a_9=\a_{10}=\a_{12}=1,\nn\\&\a_1=\a_4=\a_9=\a_{7}=-1,\nn\\&\a_2=\a_6=\a_8=\a_{11}=0,\nn\\
&Z^c_{4R}=Z^c_{3R}\;\;Z_{3T}^c=-Z^c_{4T}=Z^c_{1T}.
 \eea

%%%%%%%%%%%%%%%%%%%%%%%%%%%%%%%%%%%%%%%%%%%%%%%%%%%%%%%%%%%%%%%%%%%%%%%%%%%%%%%%%%%
\section{Conclusion}\label{conclusions}
We have demonstrated that the space of Seiberg duals to a given quiver gauge theory inherits the
action of a finite Heisenberg group. An interesting new feature is that to consider the action of the
Heisenberg group we are forced to enlarged the space from one quiver to the set of quiver gauge theories arising
from performing Seiberg duality at various nodes.

It has been shown that different Seiberg phases are related to
different toric phases \cite{seibergquiver, td1, td3}. It would, therefore, be nice to study the action of
the Heisenberg group purely as a symmetry in the space of toric phases. This is particularly interesting in the
space of quiver gauge theories which are Seiberg dual with different ranks of the gauge groups.

Assuming that there is a well-defined string theory dual for each Seiberg phase, our findings imply
that charges of branes of one string theory are related to the charges of branes of a different theory.
In a sense this provide a sort of {\it stringy toric duality}. A perhaps more speculative way of describing this situation is
that string theory views different Seiberg duals as twisted sectors of a given theory. It would be interesting to explore these 
suggestions and we hope to return to some of these questions in the future.

\section*{Acknowlegments}
We are grateful to Ami Hanany for a very enlightening discussion. BB wishes to thank
David Morrissey and Paul de Medeiros for useful conversations and clarifications. This work
is  partially supported by Department of Energy under grant
DE-FG02-95ER40899 to the University of Michigan

\end{document}